\documentclass{emulateapj}
\usepackage{apjfonts}
\usepackage{graphicx}

\usepackage{amsmath}

\shorttitle{Richtmyer-Meshkov Type Instability of a Current Sheet}
\shortauthors{T. INOUE}

\begin{document}

\title{
RICHTMYER-MESHKOV TYPE INSTABILITY OF A CURRENT SHEET IN A RELATIVISTICALLY MAGNETIZED PLASMA
}
\author{Tsuyoshi Inoue}
\altaffiltext{1}{Department of Physics and Mathematics, Aoyama Gakuin University, Sagamihara, Kanagawa 252-5258, Japan; inouety@phys.aoyama.ac.jp}

\begin{abstract}
Linear stability of a current sheet that is subject to an impulsive acceleration due to a shock passage is studied with the effect of guide magnetic field.
We find that the current sheet embedded in relativistically magnetized plasma always shows a Richtmyer-Meshkov type instability, while it depends on the density structure in the Newtonian limit.
The growth of the instability is expected to generate turbulence around the current sheet that can induce so-called turbulent reconnection whose rate is essentially free from plasma resistivity.
Thus, the instability can be applied as a triggering mechanism of rapid magnetic energy release in variety of high-energy astrophysical phenomena such as pulsar wind nebulae, gamma-ray bursts, and active galactic nuclei, where the shock wave is supposed to play a crucial role.
\end{abstract}


\section{Introduction}
Release of magnetic energy is believed to be essentially important in high-energy astrophysical phenomena such as pulsar wind nebulae, gamma-ray bursts, and active galactic nuclei (e.g., Kennel \& Coroniti 1984, Giannios et al. 2009, Lyubarsky 2010).
However, the detailed physics of the dissipation mechanism of the magnetic field is still unclear.
Theoretical studies have been suggesting that instability of current sheet or turbulent environment enhances the rate of magnetic reconnection substantially independent of plasma resistivity (Lazarian \& Vishniac 1999, Kowal et al. 2009).
Takamoto et al. (2012) also showed that turbulent stretching of current sheet leads to the dissipation of the magnetic field in turbulent eddy quickly in a few eddy turnover time regardless of plasma resistivity.

Recently, the current sheet embedded in relativistically magnetized plasma under the effect of secular acceleration was found to be unstable (Lyubarsky 2010).
Here the relativistically magnetized plasma means that the magnetic energy density is larger than rest mass energy density.
Despite the instability found by Lyubarsky is a Rayleigh-Taylor (or Kruskal-Schwarzchild) type instability, it can be unstable even when the system is isochoric, since the magnetic and the thermal energies play the role of inertia in the relativity.
The Richtmyer-Meshkov instability, which is induced by an impulsive acceleration due to shock passage (Richtmyer 1960, Nishihara et al. 2010), is a counterpart of the Rayleigh-Taylor instability, suggesting that the shock-current sheet interaction in the relativistically magnetized plasma can be unstable as a counterpart of Lyubarsky's instability.
Since the shock wave is an essential ingredient in the high-energy astrophysical phenomena, the shock-current sheet interaction is quite ubiquitous.

For these reasons, we study linear stability of the relativistic current sheet that is subject to an impulsive acceleration due to the shock passage.
The organization of this paper is as follows.
In \S 2, we provide unperturbed zeroth-order state of the current sheet.
Then, in \S 3, the linear stability analysis is performed and the master equation that governs instability growth is derived.
In \S 4, basic properties of the instability is studied based on the solution of the master equation.
Finally, in \S 5, we discuss implications of the instability.

\section{Zeroth-order State}
In this paper, we use units of the speed of light $c=1$ and the magnetic permeability $\mu=1$.
We consider a static initial current sheet as follows:
Magnetized media that have oppositely oriented $z$-component magnetic field are separated by a current sheet.
In the left and right sides of the current sheet (henceforth region 1 and region 3, respectively), the magnetic field is in $y$-$z$ plane $\vec{B}_{1(3)}=(0,B_{y,1(3)},B_{z,1(3)})$, where $B_{z,1}=-B_{z,3}$.
The $y$-component magnetic field, so called the guide field, is constant across the current sheet ($B_{y,1}=B_{y,2}=B_{y,3}\equiv B_y$).
Inside the current sheet, where the $z$-component magnetic field is dissipated, the total pressure is balanced with external plasma as $p_{2}+B_{y,2}^{2}/2=p_{1(3)}+B_{y,1(3)}^{2}/2+B_{z,1(3)}^{2}/2$.
If the plasma in region 1 (and 3) are relativistically magnetized and cold ($B_{z,1}\gg\rho_1,\,p_{1}$), the ratio of the inertia (or total enthalpy) of the media 1 and 2 is $w_{2}/w_{1}=\{\rho_2 + \gamma_2\,p_{2}/(\gamma_2-1)+B_{y,2}^2\}/\{\rho_1+\gamma_1\,p_{1}/(\gamma_1-1)+B_{y,1}^2+B_{z,1}^2\}\simeq (2\,B_{z,1}^2+3\,B_{y,1}^2)/(B_{z,1}^2+B_{y,1}^2)$, where the adiabatic index $\gamma_2=4/3$ is used in the last expression for the relativistically hot plasma of the region 2 ($\rho_2\ll p_2\sim O[B_1^2]$).
In particular, when there is no guide field, the ratio becomes $w_{2}/w_{1}\simeq2$.
This indicates that the current sheet is heavier than the external plasma, and this is the essence of the instability shown below.

Let us consider the propagation of a fast relativistic-magnetohydrodynamic (RMHD) shock wave in the region 1 toward $+x$ direction.
In the relativistically magnetized cold plasma ($B_1^2>\rho_1>p_1$), the shock velocity is close to the speed of light ($v_{\rm sh}\simeq1$), because the fast characteristic speed is close to the light velocity, while the fast shock causes only a small jump of fluid velocity (also density and magnetic field strength as well).
According to the jump condition of the RMHD shock for such plasma (Kennel \& Coroniti 1984), the post shock plasma velocity in the upstream rest-frame is given by
\begin{equation}\label{Vmax}
V\simeq \frac{v_{\rm sh}^2-2\,(1-v_{\rm sh})\,B_1^2/\rho_1}{v_{\rm sh}^3+2\,(1-v_{\rm sh})\,B_1^2/\rho_1}.
\end{equation}
By solving Eq.~(\ref{Vmax}), we see that the post shock velocity can be subsonic ($V<c_{\rm s}=1/\sqrt{3}$), provided
$B_1^2/\rho_1> v_{\rm sh}^2\,(1-c_{\rm s}\,v_{\rm sh})/\{2\,(1+c_{\rm s})\,(1-v_{\rm sh}) \}.$
Thus, for instance, even in the cases of $v_{\rm sh}=0.9,\,0.99,$ and $0.999$, the post shock velocity is subsonic for the medium of $B_{1}^2/\rho_1>1.23,\,13.3,$ and $134$, respectively.
The postshock velocity $V$ basically gives the degree of impulsive acceleration for the current sheet.
We discuss about this point in more detail in the next section.

\section{Linear Stability Analysis}
\begin{figure}[t]
\epsscale{1.}
\plotone{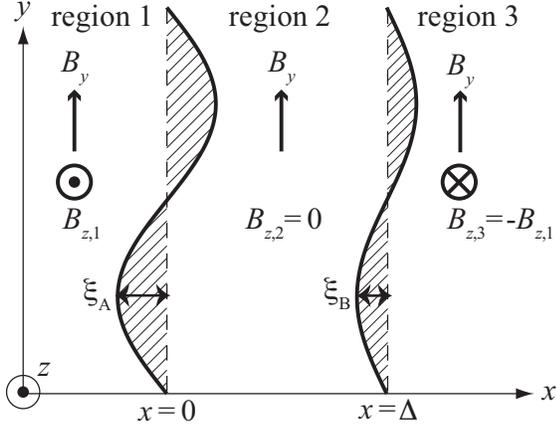}
\caption{
Schematic of the system under consideration.
Shaded areas are the regions for the perturbations that subject to the acceleration compared to the zeroth order state.
}
\label{f1}
\end{figure}

If the initial current sheet is uniform, the shocked media are accelerated and move homogeneously after the shock passage.
However, when the current sheet is perturbed as illustrated in Fig.~\ref{f1}, the shock acceleration near the current sheet acts inhomogeneously depending on the $y$-coordinate due to the different inertia of the media.
In the following, we examine the dynamics of perturbations in the rest-frame of the homogeneous shocked current sheet, indicating that the zeroth order variables are taken as their postshock values when the current sheet is uniform.
We denote the left (right) side of the interface by the interface A (B), its amplitude by $\xi_{\rm A}$ ($\xi_{\rm B}$), and the mean position by $x=0$ ($x=\Delta$).

We consider perturbation due to an impulsive shock passage that is imposed as a force term  on the basic equations given below following the formulation by Wheatley et al. (2005), and the linearization is performed in order to find the response of such a perturbation.
Since the perturbed flow can be treated as subsonic one owing to eq. (\ref{Vmax}), we can assume the Lorentz factor of the perturbed flows to be unity and employ the incompressive approximation.
Then, if we write the perturbed variables by using the small letters and the unperturbed zeroth order variables by the capital letter, the basic RMHD equations for the perturbations can be written as
\begin{eqnarray}
&&\partial _x\,v_x+\partial _y\,v_y=0,\label{p1}\\
&&(W\!\!+\!B_y^2\!+\!B_z^2)\partial_t v_x\!=\!-\partial_x p\!-\!B_y \partial_x b_y\!\!+\!B_y\partial_y b_x\!\!+\!F(t,x,y),\label{p2}\\
&&(W+B_z^2)\,\partial_t\,v_y=-\partial_y\,p,\label{p3}\\
&&\partial_t\,b_x=B_y\,\partial_y\,v_x,\label{p4}\\
&&\partial_t\,b_y=-B_y\,\partial_x\,v_x,\label{p5}\\
&&\partial_x\,b_x+\partial_y\,b_y=0,\label{p6}
\end{eqnarray}
where $W=\rho+\gamma\,P/(\gamma-1)$ is the specific enthalpy and $F(t,x,y)$ is the force term that represents the acceleration of the perturbed flow.
In the above expression, we have omitted the $z$-component velocity and magnetic field perturbations owing to the fact that they are decoupled from the other variables and only describe the propagation of the Alfv\'en wave in $y$-direction.

Because the shock crossing time of the current sheet can be much smaller than the timescale of perturbations, the force term can be modeled as
\begin{eqnarray}
F(t,x,y)&=&(W_2+B_{y,2}^2+B_{z,2}^2-W_1-B_{y,1}^2-B_{z,1}^2) \nonumber\\
&&\times\,\{ H[x]-H[x-\xi_{\rm A}(y,t)] \}\,V\,\delta(t) \nonumber\\
&&+(W_3+B_{y,3}^2+B_{z,3}^2-W_2-B_{y,2}^2-B_{z,2}^2) \nonumber\\
&&\times\,\{ H[x+\Delta]-H[x+\Delta-\xi_{\rm B}(y,t)] \}\,V\,\delta(t),\label{IA}
\end{eqnarray}
where $H(t)$ is the Heaviside function and $\delta(t)$ is the Dirac delta function (see also Wheatley et al. 2005).
The regions for the perturbations that subject to the acceleration compared to the zeroth order state are illustrated as shaded regions in Fig.~\ref{f1}.
In this formulation, we have to careful about following points:
In Wheatley et al. (2005) that studied the Richtmyer-Meshkov instability (instability of a single contact surface), the speed of contact surface after the shock passage is chosen as $V$ in eq. (\ref{IA}), which is slightly different from the postshock velocity.
Thus, if we consider the case where there is only the interface A, the speed of the contact (interface A) between the shocked regions 1 and 2 must be used as $V$ in eq. (\ref{IA}), which is obtained by solving an appropriate Riemann problem (see, e.g., Giacomazzo \& Rezzolla 2006).
This suggests that for the stability of the interface A in the short-wavelength limit (see eq. [\ref{eA}] below), the substitution of the speed of the shocked contact as $V$ in eq. (\ref{IA}) rather than the postshock velocity of eq. (\ref{Vmax}) would give more accurate growth rate.
For the finite thickness current sheet, the speed of the interfaces A and B after the shock passage approaches the postshock velocity $V$ of eq. (\ref{Vmax}) asymptotically with time.
Thus, since the shock crossing time of the current sheet is much smaller than growth timescale of the instability, the post shock velocity of eq. (\ref{Vmax}) is more appropriate as the degree of the impulsive acceleration $V$ in eq. (\ref{IA}).
Another point we have to keep in mind is the neglect of the effect of a reflection shock (or a rarefaction wave) that is generated when the incident shock hits the interface B, which causes complex dynamics in the current sheet.
However, direct numerical simulations of the (Newtonian and unmagnetized) Richtmyer-Meshkov instability of a finite thickness fluid layer showed that the impulsive model can indeed work well, although it cause error about a factor 2 or less in growth rate (Mikaelian 1996).

Assuming the perturbations of the form $q(x,y,t)=\hat{q}(x,t)\,\exp(i\,k\,y)$ and taking the temporal Laplace transform ($\bar{q}[x,s]=\int_{0}^{\infty} \hat{q}[x,t]\,\exp[-s\,t]\,dt$) of (\ref{p1})-(\ref{p6}) outside of the forced region give
\begin{equation}
\partial _x\,\bar{v}_x+i\,k\,\bar{v}_y=0,\label{pp1}
\end{equation}
\begin{equation}
s\,(W+B_y^2+B_z^2)\,\bar{v}_x=-\partial_x\,\bar{p}-B_y\,\partial_x\,\bar{b}_y+i\,k\,B_y\,\bar{b}_x,\label{pp2}
\end{equation}
\begin{equation}
s\,(W+B_z^2)\,\bar{v}_y=-i\,k\,\bar{p},\label{pp3}
\end{equation}
\begin{equation}
s\,\bar{b}_x=i\,k\,B_y\,\bar{v}_x,\label{pp4}
\end{equation}
\begin{equation}
s\,\bar{b}_y=-B_y\,\partial_x\,\bar{v}_x,\label{pp5}
\end{equation}
\begin{equation}
\partial_x\,\bar{b}_x+i\,k\,\bar{b}_y=0,\label{pp6}
\end{equation}
where $k$ is the wavenumber of perturbation and $s$ is the variable associated with the temporal Laplace transformation.
Note that these equations describe the perturbations outside the forced regions.
The effect of the force term is taken into account when the perturbations in the different regions are connected across the forced regions (when deriving Eqs.~[\ref{pb1}] and [\ref{pb2}]).
Eliminating $\bar{p}$, $\bar{b}_x$, and $\bar{b}_y$ in Eq.~(\ref{pp2}) by using Eqs.~(\ref{pp3})-(\ref{pp5}) and (\ref{pp1}), we obtain the following ordinary differential equation for $\bar{v}_x$:
\begin{eqnarray}
&&\left(\phi\,D_x^2-\chi\,k^2 \right)\,\bar{v}_x=0,\label{ppp}\\
&&\phi=B_y^2\,k^2+s^2\,(B_z^2+W),\nonumber\\
&&\chi=B_y^2\,(k^2+s^2)+s^2\,(B_z^2+W),\nonumber
\end{eqnarray}
which has the general solution of the form
\begin{equation}
\bar{v}_x=\alpha\,\exp(\sqrt{\chi/\phi}\,k\,x)+\beta\,\exp(-\sqrt{\chi/\phi}\,k\,x).
\end{equation}
Other variables are expressed by using $\bar{v}_x$, e.g.,
\begin{eqnarray}
&&\bar{p}=-s\,(W+B_z^2)\partial_x\,\bar{v}_x/k^2,\label{ppp}\\
&&\bar{b}_{y}=-B_y\,\partial_x\,\bar{v}_x/s.\label{pby}
\end{eqnarray}
Because the perturbations must vanish at $x=\pm\infty$, $\beta_1$ ($\beta$ in region 1) and $\alpha_3$ ($\alpha$ in region 1) should be null coefficients.
We write $\bar{v}_{x}$ in each region as follows: 
\begin{eqnarray}
\bar{v}_{x,1}&=&\alpha_1\,e^{\sqrt{\chi/\phi}\,k\,x},\label{pvx1}\\
\bar{v}_{x,2}&=&\alpha_2\,e^{\sqrt{\chi/\phi}\,k\,(x-\Delta)}+\beta_2\,e^{-\sqrt{\chi/\phi}\,k\,x},\label{pvx2}\\
\bar{v}_{x,3}&=&\beta_3\,e^{-\sqrt{\chi/\phi}\,k\,(x-\Delta)}\label{pvx3}.
\end{eqnarray}

Let us consider junction conditions for perturbations at the interfaces.
At the moment, we have four undetermined amplitude of perturbations $\alpha_1,\,\alpha_2,\,\beta_2$, and $\beta_3$, indicating that total four junction conditions are necessary to determine them.
Two of them are the continuity of the velocity normal to the interfaces that leads to $\bar{v}_{x,1}(0,s)=\bar{v}_{x,2}(0,s)$ and $\bar{v}_{x,2}(\Delta,s)=\bar{v}_{x,3}(\Delta,s)$ to the first order of the perturbed variables.
Substituting (\ref{pvx1})-(\ref{pvx3}) to the conditions, we obtain
\begin{eqnarray}
&&\alpha_1=\alpha_2\,\exp(-\sqrt{\chi/\phi}\,k\,\Delta)+\beta_2,\label{c1}\\
&&\alpha_2+\beta_2\,\exp(-\sqrt{\chi/\phi}\,k\,\Delta)=\beta_3.\label{c2}
\end{eqnarray}
The remaining two conditions, which describe the force balance at each interface, are obtained by integrating (\ref{p2}) with regard to $z$ across each inhomogeneous region.
After the temporal Laplace transforming of the integrated equations, to the first order of the perturbed variables, we obtain
\begin{eqnarray}
&&(\bar{p}_2[0,s]+B_{y,2}\,\bar{b}_{y,2}[0,s])-(\bar{p}_1[0,s]+B_{y,1}\,\bar{b}_{y,1}[0,s])\nonumber\\
&&\ \ \ \ \ \ \ \ \ \ \ \ \ \ =(W_{{\rm tot},2}-W_{{\rm tot},1})\,V\,\hat{\xi}_{\rm A,0},\label{pb1}\\
&&(\bar{p}_3[\Delta,s]+B_{y,3}\,\bar{b}_{y,3}[\Delta,s])-(\bar{p}_2[\Delta,s]-B_{y,2}\,\bar{b}_{y,2}[\Delta,s])\nonumber\\
&&\ \ \ \ \ \ \ \ \ \ \ \ \ \ =(W_{{\rm tot},3}-W_{{\rm tot},2})\,V\,\hat{\xi}_{\rm B,0},\label{pb2}
\end{eqnarray}
where $\hat{\xi}_{0}$ is the initial amplitude of the interface, and we have defined the total enthalpy $W_{\rm tot}\equiv W+B_y^2+B_z^2$.
In the above expression, we have used the fact that $\bar{v}_{x}$ is continuous across the interface and also the case of $\bar{b}_{x}$ due to (\ref{pp4}).
Substituting (\ref{pvx1})-(\ref{pvx3}) into (\ref{pb1}) and (\ref{pb2}) via (\ref{ppp}) and (\ref{pby}), we obtain two conditions written by $\alpha_1,\alpha_2,\beta_2,$ and $\beta_3$.
Solving these two conditions, (\ref{c1}), and (\ref{c2}) with respect to the four coefficients, we get
\begin{eqnarray}
&&\alpha_1 = \frac{A\,\hat{\xi}_{\rm A,0}+B\,\hat{\xi}_{\rm B,0}}{C}\,k\,s\,V\,(W_{{\rm tot},2}-W_{{\rm tot},1}),\label{amp1}\\
&&\beta_3 = -\frac{A\,\hat{\xi}_{\rm B,0}+B\,\hat{\xi}_{\rm A,0}}{C}\,k\,s\,V\,(W_{{\rm tot},2}-W_{{\rm tot},1}),\label{amp2}\\
&&A=\!  \{ W_{{\rm tot},2}\!+\!W_{{\rm tot},1}\!-\!2\,B_y^2\!+\!(W_{{\rm tot},2}\!-\!W_{{\rm tot},1})\,e^{-2\,k\,\Delta} \!\} \,s^2 \nonumber\\
&&\ \ \ \ \ \ \ +2\,B_y^2\,k^2,\nonumber\\
&&B=\! \{ 2\,(W_{{\rm tot},2}-B_y^2)\,s^2+2\,B_y^2\,k^2 \}\, e^{-k\,\Delta},\nonumber\\
&&C=\!  \{ (W_{{\rm tot},2}+W_{{\rm tot},1}-2\,B_y^2)\,s^2+2\,B_y^2\,k^2 \}^2 \nonumber\\
&&\ \ \ \ \ \ \ \ -(W_{{\rm tot},2}-W_{{\rm tot},1})^2\,e^{-2\,k\,\Delta} \,s^4,\nonumber
\end{eqnarray}
where we have used the zeroth order conditions $W_{{\rm tot},3}=W_{{\rm tot},1}$ and $B_y\equiv B_{y,1}=B_{y,2}=B_{y,3}$, and again assumed slow motion of the perturbed flows compared to the speed of light: $\chi /\phi\simeq1$.

The coefficients $\alpha_1$ and $\beta_3$ are equivalent to the Laplace transform of the interface velocities ${\cal L}[\partial_t\,\hat{\xi}_{\rm A}(t)]$ and ${\cal L}[\partial_t\,\hat{\xi}_{\rm B}(t)]$, respectively.
The inverse Laplace transform of (\ref{amp1}) and (\ref{amp2}) yield the temporal differential equations for the two interfaces:
\begin{eqnarray}
&&\frac{d}{dt}\,\hat{\xi}_{\rm A}(t)= (W_{{\rm tot},2}-W_{{\rm tot},1})\,k\,V\,H(t)\nonumber\\
&&\ \ \ \times \Big[\,\mu\,(1+e^{-k\,\Delta})(\hat{\xi}_{\rm A,0}-\hat{\xi}_{\rm B,0})\cos(2\,\sqrt{\mu}\,B_y\,k\,t) \nonumber\\
&&\ \ \ \ \ +\nu\,(1-e^{-k\,\Delta})(\hat{\xi}_{\rm A,0}+\hat{\xi}_{\rm B,0})\cos(2\,\sqrt{\nu}\,B_y\,k\,t) \,\Big],\label{eqAB}\\
&&\mu^{-1}=2\{ W_{{\rm tot},2}(1-e^{-k\,\Delta})+W_{{\rm tot},1}(1+e^{-k\,\Delta})-2\,B_y^2 \},\nonumber\\
&&\nu^{-1}=2\{ W_{{\rm tot},2}(1+e^{-k\,\Delta})+W_{{\rm tot},1}(1-e^{-k\,\Delta})-2\,B_y^2 \},\nonumber
\end{eqnarray}
where the equation for the interface B is obtained by changing the overall sign of the right hand side of (\ref{eqAB}) and by interchanging the subscripts A and B.

\section{Properties of Solution}
\subsection{Short wave-length limit}
Let us first consider small-scale perturbations compared to the thickness of the current sheet, which leads to independent evolutionary equations of the two interfaces.
Taking the limit of $k\,\Delta\rightarrow \infty$ in (\ref{eqAB}) and integrating with regard to $t$, we obtain
\begin{eqnarray}
\hat{\xi}_{\rm A,B}(t)&=&\hat{\xi}_{\rm A,B}(0)\pm {\cal A}\,V\,k\,\tau\,\hat{\xi}_{\rm A,B}(0)\,\sin\left( t/\tau \right), \label{eA}\\
{\cal A}&=&\frac{W_{{\rm tot},2}-W_{{\rm tot},1}}{W_{{\rm tot},2}+W_{{\rm tot},1}-2\,B_y^2},\label{AWN}\\
\tau&=& (W_{{\rm tot},2}+W_{{\rm tot},1}-2\,B_y^2 )^{1/2}/ (\sqrt{2}\,B_y\,k).
\end{eqnarray}
where plus (minus) sign in the second term of the right hand side of (\ref{eA}) is for the interface A (B), and $\tau$ represents the lateral Alfv\'en crossing time.
Eq. (\ref{eA}) shows that, as long as $t\ll\tau$ (and ${\cal A}\ne0$), interface deformation grows linearly with time with the velocity of
\begin{equation}\label{RM}
\dot{\xi}={\cal A}\,V\,k\,\hat{\xi}(0)\,e^{i\,k\,y}.
\end{equation}
For $t>\tau$, the interface oscillates at the Alfv\'en frequency $\omega\sim B_y\,k/W_{\rm tot}^{1/2}$ due to the magnetic tension force.
In the case where the guide field is absent ($B_y=0$), the interface grows perpetually with this speed.
In the Newtonian limit ($\rho\gg P$ and $B^2$), ${\cal A}$ reduces to the Atwood number ${\cal A}=(\rho_2-\rho_1)/(\rho_2+\rho_1)$, and the growth velocity (\ref{RM}) exactly recover that of the original Richtmyer-Meshkov instability (Richtmyer 1960).
In addition, when $B_y=0$, the condition ${\cal A}>0$ reduces to the criterion of Lyubarsky's instability.
Thus, the coefficient ${\cal A}$ can be regarded as the generalized Atwood number.
It is noteworthy that, differently from the Rayleigh-Taylor type instability (including Lyubarsky's instability), the Richtmyer-Meshkov type instability can grow even when the Atwood number is negative, i.e. the interface B is also unstable.
The generalized Atwood number of the current sheet in the relativistically magnetized plasma ($B_{z,1}^2\gg W_1$ and $P_{2}\gg \rho_2$) is ${\cal A}=1/3$, indicating that the interface is always unstable with respect to the shock passage.
On the other hand, in the Newtonian limit, the current sheet can be unstable provided its density is different from the external plasma (in the case of the isothermal Harris current sheet, the Atwood number is ${\cal A}=(1+2\,\beta)^{-1}$, where $\beta$ is the ratio of the thermal to magnetic pressure in the region 1).

\subsection{Long wave-length limit}
\begin{figure}[t]
\epsscale{1.}
\plotone{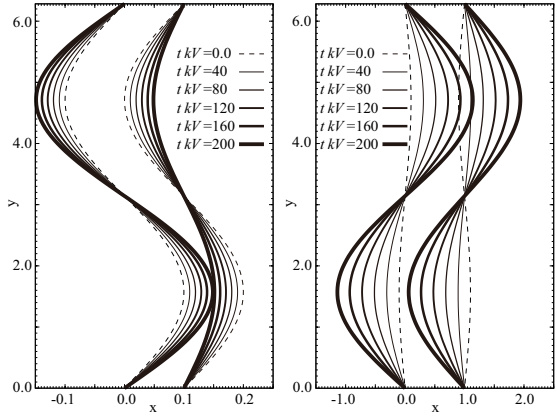}
\caption{
Evolutionally sequence of the solution (\ref{eqAB}).
Left panel is the case of $k\Delta=0.1$ and $\hat{\xi}_{\rm A}/\hat{\xi}_{\rm B}=1$, and right panel is the case of $k\Delta=1$ and $\hat{\xi}_{\rm A}/\hat{\xi}_{\rm B}=-1$.
Other parameters are $B_y=0,\,W_{{\rm tot},2}=2,\,W_{{\rm tot},1}=1$ and $\hat{\xi}_{\rm B}=0.1$.
}
\label{f2}
\end{figure}

Next, we consider large-scale perturbations compared to the thickness of the current sheet ($k\,\Delta\ll1$).
The Taylor expansion of (\ref{eqAB}) with regard to $k\,\Delta$ results in
\begin{eqnarray}
\frac{d}{dt}\,\hat{\xi}_{\rm A}(t)&=&-\frac{d}{dt}\,\hat{\xi}_{\rm B}(t)\label{eqABS1}\\
&=&\! \frac{W_{{\rm tot},2}\!-\!W_{{\rm tot},1}}{W_{{\rm tot},1}-B_y^2}Vk^2\!\Delta\,\hat{\xi}_{\rm A,0}\cos(t/\tau\!)H(t), \label{eqABS2} \\
\tau&=&(W_{{\rm tot},1}-B_y^2 )^{1/2}/ (B_y\,k),\nonumber
\end{eqnarray}
where we have used the fact that the initial amplitudes $\hat{\xi}_{\rm A,0}$ and $\hat{\xi}_{\rm B,0}$ must be equal in this limit.
Eq. (\ref{eqABS1}) shows that the two interfaces move toward opposite directions.
The evolutionally sequence of the interfaces for the long wavelength perturbation is shown in left panel of Fig.~\ref{f2}.
The growth in opposite phase is not surprising, because the generalized Atwood number ${\cal A}$ at the two interfaces have opposite signs.

\subsection{General properties}
In the general case including $k\,\Delta\sim1$, as the cases in both short and long wavelength limits, the following items are true:
(i) The interfaces can be unstable if the generalized Atwood number ${\cal A}$ is nonzero.
(ii) In $t\ll\tau$ or $B_y=0$, where $\tau\sim\sqrt{W_{\rm tot}}/(B_y\,k)$ is the lateral Alfv\'en crossing time, the interfaces grow with constant speed of $\sim {\cal A}\,k\,V\,\xi_0$.
(iii) In $t>\tau$, the interfaces oscillate with the frequency $\sim 1/\tau$.
Interestingly, eq. (\ref{eqAB}) suggests that, for $B_y=0$ (or $t\ll\tau$), the growth speed of the interface A ($d\xi_{\rm A}/dt$) becomes zero for the special case of $\hat{\xi}_{\rm A,0}/\hat{\xi}_{\rm B,0}=\{(W_{\rm tot,2}+W_{\rm tot,1})+(W_{\rm tot,2}-W_{\rm tot,1})\,e^{-2k\,\Delta}\}/(2W_{\rm tot,2}\,e^{-k\,\Delta})\equiv r_{\rm c}$ and that of the interface B becomes zero for $\hat{\xi}_{\rm A,0}/\hat{\xi}_{\rm B,0}=r_{\rm c}^{-1}$, where $r_{\rm c}$ is always larger than unity.
Furthermore, when the ratio of the initial amplitudes is in the range $r_{\rm c}>\hat{\xi}_{\rm A,0}/\hat{\xi}_{\rm B,0}>r_{\rm c}^{-1}$, the ratio of the growth speeds of the interface A and B is negative (out of phase growth).
When the initial ratio is larger than $r_{\rm c}$ or smaller than $r_{\rm c}^{-1}$ including negative value, the ratio of the growth speeds is positive (in phase growth).
The reason for this as follows:
Since the generalized Atwood number ${\cal A}$ has opposite signs at the two interfaces, the interfaces at the same $y$ are accelerated toward the opposite (same) directions when the initial interfaces are in phase (out of phase).
However, when the magnitude of the initial amplitude is very different, the motion of the interface with minor initial amplitude is dragged toward major one, because the volume of the accelerated regions depends on the initial amplitude (see Fig.~\ref{f1}).
The evolutionally sequences of the in phase and the out of phase growths are shown in the left and right panels of Fig~\ref{f2}, respectively.

\section{Discussion}
Finally, we discuss implications of the instability.
Recent particle-in-cell simulation of a relativistically magnetized pulsar wind have shown that shock propagation through the stripes of opposite magnetic field polarity induces driven magnetic reconnection (Sironi \& Spitokovsky 2011).
The triggering mechanism of the magnetic reconnection could be the instability studied in this article.

In addition, recent MHD simulations have shown that the growth of Richtmyer-Meshkov instability induced by a (relativistic) shock propagation through inhomogeneous density medium generates turbulence in its nonlinear stage (Giacalone \& Jokipii 2007, Inoue et al. 2009, 2011, 2012, Beresnyak et al. 2009).
Because the instability of the relativistic current sheet found in this study is also a Richtmyer-Meshkov type instability, it is quite reasonable to expect the development of turbulence.

The instability evolves into the nonlinear stage, if the amplitude of the interface can grow to $\xi\sim 1/k$ before the magnetic tension force begins to suppress the growth ($t\lesssim\tau$).
Using the growth velocity (\ref{RM}), the above condition is reduced to
\begin{equation}
\frac{V}{V_{{\rm A},y}}\gtrsim ({\cal A}\,k\,\xi_0)^{-1},\label{NL}
\end{equation}
where $V_{{\rm A},y}\equiv B_y/\sqrt{W_{\rm tot}}$ is the lateral Alfv\'en velocity.
Note that if there is no guide field ($B_y=0$), which is plausible at least in the pulsar wind, the instability can always go into its nonlinear stage.
As in the case of the Richtmyer-Meshkov instability, the growth of the instability is not exponential with time but linearly with time (Richtmyer 1960), indicating that the timescale of the evolution depends on the initial amplitude of perturbation.
If we consider the initial perturbation of scale $k\gtrsim \Delta^{-1}$ with amplitude $\xi_{{\rm B},0}=\xi_{{\rm A},0}=\xi_0$ in which the instability is the most influential, the instability grows to nonlinear ($\xi\,k\sim 1$) after the following elapsed time from the shock passage:
\begin{equation}
t\sim ({\cal A}\,V\,\xi_0\,k^2)^{-1},
\end{equation}
where we have estimated the timescale based on eq (\ref{eqAB}).
This suggests that, if the initial amplitude is much smaller than the wavelength of perturbation ($\lambda=k^{-1}\ll \xi_0$), it takes huge time for the instability to evolve into the nonlinear stage.
Fortunately, however, we can expect large initial amplitude of perturbation owing to the tearing mode instability and/or Lyubarsky's instability that can grow prior to the shock passage.
It is discussed in Lyubarski (2011) that Lyubarsky's instability can grow in pulsar winds, gamma-ray burst jets, and active galactic nuclei at least in the scales $\lambda\lesssim \Delta $, although it may not fully dissipate the magnetic field especially in pulsar winds.
Thus, we can expect excitation of turbulence quickly (on the order of crossing time $\lambda\,V^{-1}$ for $\xi_0\sim \lambda$) after the shock passage.
Note that, in Lyubarsky's instability, either the interface A or B can be unstable depending on the orientation of acceleration in the most unstable scales of $\lambda \lesssim \Delta $, while our instability can grow in both interfaces.
Therefore, even if Lyubarsky's instability is going on, our instability that agitates the current sheet as a hole is necessary in order to excite "turbulence".

Once the current sheet becomes turbulent, the induction of so called ``turbulent reconnection" (Lazarian \& Vishniac 1999, Kowal et al. 2009) and the effect of turbulent stretching of the current sheet (Takamoto et al. 2012) dissipate the magnetic field rapidly regardless of plasma conductivity.
The excited turbulent reconnection would then evolve through generation of more turbulence by the positive feedback of reconnection outflows.
Therefore, the instability found in this study can be applied as a triggering mechanism of rapid magnetic energy release in variety of high-energy astrophysical phenomena such as pulsar wind nebulae, gamma-ray bursts, and active galactic nuclei, where the shock wave is supposed to play a crucial role.
Differently from the secular instabilities such as the tearing mode instability and Lyubarsky's instability, sudden onset of our instability due to the shock passage and following turbulent reconnection may be preferred especially in the intermittent phenomena such as gamma-ray bursts.

T.I. is grateful for S. Inutsuka, T. Sano, K. Ioka, R. Yamazaki, Y. Ohira and M. Takamoto.
This work could not be started without the discussions with them.
This work is supported by Grant-in-aids from the Ministry of Education, Culture, Sports, Science, and Technology (MEXT) of Japan, No.22$\cdot$3369 and No. 23740154.

\end{document}